\documentstyle[12pt]{article}
\newcommand{\p}[1]{(\ref{#1})}
\def\PRD#1#2#3{{\sl Phys. Rev.} {\bf D#1} (#2) #3}

\def\PLB#1#2#3{{\sl Phys. Lett.} {\bf #1B} (#2) #3}

\def\LMP#1#2#3{{\sl Letters in Math. Phys.} {\bf #1} (#2) #3}

\def\TMP#1#2#3{{\sl Theor. Math. Phys.} {\bf #1} (#2) #3}

\def\JETPL#1#2#3{{\sl  Sov. Phys. JETP Lett.} {\bf #1} (#2) #3}

\def\CQG#1#2#3{{\sl Class. Quantum Grav.} {\bf #1} (#2) #3}

\def\Ai{\hbox{\hbox{${\cal A}$}}\kern-1.9mm{\hbox{${/}$}}}
\def\Vi{\hbox{\hbox{${\cal V}$}}\kern-1.9mm{\hbox{${/}$}}}
\def\Di{\hbox{\hbox{${\cal D}$}}\kern-1.9mm{\hbox{${/}$}}}
\def\lam{\hbox{\hbox{${\lambda}$}}\kern-1.6mm{\hbox{${/}$}}}
\def\D{\hbox{\hbox{${D}$}}\kern-1.9mm{\hbox{${/}$}}}
\def\A{\hbox{\hbox{${A}$}}\kern-1.8mm{\hbox{${/}$}}}
\def\V{\hbox{\hbox{${V}$}}\kern-1.9mm{\hbox{${/}$}}}
\def\parz{\hbox{\hbox{${\partial}$}}\kern-1.7mm{\hbox{${/}$}}}
\def\B{\hbox{\hbox{${B}$}}\kern-1.7mm{\hbox{${/}$}}}
\def\R{\hbox{\hbox{${R}$}}\kern-1.7mm{\hbox{${/}$}}}
\def\si{\hbox{\hbox{${\xi}$}}\kern-1.7mm{\hbox{${/}$}}}

\begin{document}
\thispagestyle{empty}
\renewcommand{\thefootnote}{\dagger}

\centerline
{\bf ON THE CHIRAL FERMIONS IN THE TWISTOR--LIKE FORMULATION\\}
\centerline
{\bf  OF D=10 HETEROTIC STRING}

\vskip 2truecm

\centerline
{\bf Dmitrij P. Sorokin\footnote{Supported in part by the European Community
Research Program ``Gauge Theories, Applied Supersymmetry and Quantum
Gravity'' under contract CEE-SCI-CT92-0789 and by M.P.I.}
\renewcommand{\thefootnote}{\ddagger}
\footnote{Permanent address: Kharkov Institute of Physics and
Technology, Kharkov, 310108, the Ukraine
e-mail address:
kfti\%kfti.kharkov.ua@relay.ussr.eu.net}
\hskip12pt and \hskip12pt Mario Tonin$^\dagger$}

\bigskip
\centerline{\sl Dipartimento di Fisica ``G. Galilei" -- Universit\`a di Padova}
\centerline{\sl Istituto Nazionale di Fisica Nucleare -- Sezione di
Padova}
\centerline{\sl Padova, Italy}

\vskip 2.5truecm

\noindent
DFPD/93/TH/52\\
July 1993\\
hep-th/

\vskip 3.5truecm

\noindent
{\bf Abstract.} An $n=8$ worldsheet superfield action is proposed for
describing chiral fermions in the twistor--like formulation of an $N=1$,
$D=10$ heterotic superstring.

\setcounter{page}1
\renewcommand{\thefootnote}{\arabic{footnote}}
\setcounter{footnote}0
\newpage

The problem of the origin of the fermionic $\kappa$--symmetry and the
related problem of the Lorentz covariant quantization of superparticle
and superstring theories (see \cite{gsw} and references therein)
stimulated the construction of versions of these theories which would
elucidate the nature of the $\kappa$--symmetry and provide a way of
covariant quantization.

In Ref. \cite{stv} it was proposed to consider the $n=N(D-2)$ parameter
$\kappa$--symmetry of the $N=1,2$ superparticles and superstrings in
$D=(2),3,4,6$ and $10$ dimensional space-time as a manifestation of
an $n$--extended worldsheet supersymmetry.

The important feature of the doubly supersymmetric approach is that the
models of this kind naturally incorporate commuting (``twistor-like'')
spinors as superpartners of the Grassmann coordinates of the
superparticles and superstrings, thus providing a ground for unifying
such fundamental trends of modern theoretical physics as supersymmetry
\cite{susy} and the twistor program \cite{tp,lli}. The relationship between
the doubly supersymmetric \cite{stv} and the supertwistor \cite{stw}
formulation of superparticle dynamics was demonstrated in \cite{gus}.

Note that a Lorentz-harmonic formulation of superparticles and
superstrings \cite{harm} developed in parallel to the twistor-like
doubly supersymmetric formulation is closely related to the latter on
the component level.

The twistor-like approach was developed in application
to superparticles \cite{spa} and superstrings \cite{sstr} by several groups,
and different aspects of the approach were discussed in
Refs.~\cite{asp}. This resulted in the construction of an $n=8$ worldline
superfield version of the $N=1,~D=10$ superparticle \cite{n8p} and
$n=(0,8)$ worldsheet superfield version of the $N=1,~D=10$ heterotic
superstring \cite{n8s,dg,bstv}. Moreover an $n=(0,2),~D=10$ heterotic
superstring
model was proved to be a way for a consistent quantization
of the $N=1,~D=10$ heterotic superstring with vanishing conformal anomaly
\cite{nbq}. Recently the n=8 superconformal
algebra has been observed in the light-cone gauge of the ordinary
formulation of the heterotic string \cite{n8}.  The twistor-like
$n=(1,1)$, $N=2$, $D=3$ Green--Schwarz
superstring formulation has been proposed in \cite{gs2}.

But the inclusion of chiral fermions into the twistor-like heterotic
string action has remained an unsolved problem.

Chiral fermions are necessary for the heterotic string theory to be
consistent at the quantum level \cite{gsw}, so their incorporation into
an $n=8$ worldsheet superfield action is important for accomplishing the
construction of the classical twistor-like heterotic string model and
may give a clue to a covariant quantization of the theory.

In the present paper we consider the problem of describing chiral fermions in
the twistor-like approach.

$D=10$, $n=8$ twistor-like heterotic string action consists of two
terms.\footnote{The reader may find the detailed analysis of the action
in Ref.~\cite{dg,asp,bstv}.} One of them is a Wess--Zumino--type term,
which serves as a string tension generation term \cite{t,dg,bstv}:
\begin{equation}\label{1}
S_{WZ}=\int d^2\xi d^8\eta {\rm P}^{\cal{MN}}\left(\hat B_{\cal{MN}}-
\partial_{\cal M}Q_{\cal N}\right),
\end{equation}
where $\xi^\pm$ and $\eta^q$ ($q=1,...,8$), (${\cal{MN}}=(+,-,q$) are,
respectively even and odd coordinates of the super worldsheet;
${\rm P}^{\cal{MN}}=(-1)^{{\cal{MN}}+1}{\rm P}^{\cal{NM}}$ is a Lagrange
multiplier superfield; $Q_{\cal N}$ denotes the components of the
pull-back of a target-superspace one-form $Q(X^m,\Theta^\mu)$ onto the
super worldsheet ($(X^m(\xi,\eta),\Theta^\mu(\xi,\eta)$ ($m=0,1,...,9;
\mu=1,...,16$) are vector and spinor coordinates of $N=1$, $D=10$ target
superspace, their dependence on $\xi,\eta$ defines the embedding of the
worldsheet onto the target space), and $\hat B_{\cal{MN}}$ are the
components of the pull-back onto the worldsheet of a two-superform
\begin{equation}\label{2}
\hat B=B+{1\over 8}e^+\wedge e^-\sum_{q=1}^{8}E^A_qE^B_qE^C_+H_{CBA}.
\end{equation}
The supervielbein one-forms $E^A$ and the two-superform $B$ (with
$H=dB$) determine the geometry of the target superspace together with
SUGRA--SYM connections $\Omega^B_A(X,\Theta)$ and $A(X,\Theta)$. This
geometry is characterized by constraints on the intrinsic components of
the background torsion and curvatures \cite{con,lli}. Capital letters $A,B,C$
from the beginning of the alphabet denote indices corresponding to the
superspace tangent to the target space, and $A=(a,\alpha)$, where $a$
denotes vector and $\alpha$ denotes spinor indices. $e^{\cal A}({\cal
A}=+,-,q)$ are the supervielbeins on the worldsheet, the geometry of the
latter being characterized by the constraints on the torsion $T^{\cal
A}=\Delta e^{\cal A}$ ($\Delta$ is a covariant differential containing
Lorentz and $SO(8)$ connection)\cite{n8s}:
\begin{equation}\label{3}
T^-=\sum_{q=1}^{8}e^q\wedge e^q;\qquad T^+=0;\qquad T^q=e^+\wedge e^-
T^q_{+-}.
\end{equation}
$E^A_{\cal A}$ are the intrinsic components of the pull-back of $E^A$.

Varying \p{1} over ${\rm P}^{\cal{MN}}$ we get $\hat
B_{\cal{MN}}=\partial_{[\cal M}Q_{\cal N)}$ ([,] and (,) denote
antisymmetrization and symmetrization of indices, respectively) from
which it follows that the pull-back of \p{2} onto the worldsheet
$M_{ws}$ must be a closed superform, that is:
\begin{equation}\label{4}
d\hat B\vert_{M_{ws}}=0.
\end{equation}

To get the consistency condition \p{4} one must take into account
constraints on the background superfields \cite{con,dg}, eqs.~\p{3} and a
twistor constraint
\begin{equation}\label{5}
E^a_q(X,\Theta)=0,
\end{equation}
which determines the embedding of the super worldsheet into the target
superspace in such a way that the odd part of the super worldsheet lies
entirely within the odd part of the target space. A consequence of \p{5}
is $\delta_{qr}E^a_-=E_q\gamma^a E_r$, which in flat target
superspace is reduced
to
\begin{equation}\label{6}
\delta_{qr}\Pi^m_-\equiv\delta_{qr}(\partial_-X^m-i\partial\Theta\gamma^m\Theta)
\vert_{\eta=0}=\lambda_q\gamma^m\lambda_r,
\end{equation}
where $\lambda_q\equiv\partial_q\Theta\vert_{\eta=0}$ are commuting
twistor-like spinors; and from eq.~\p{6} one gets a Virasoro condition
$\Pi^m_-\Pi_{m-}=0$.

The closure of $\hat B$ (eq.~\p{4}) on the twistor constraint surface \p{5},
\p{6} is nothing but a manifestation of the light-like integrability of
the $N=1$, $D=10$ SUGRA-SYM connections along a supersurface swept by
propagating superparticles and superstrings. This underlies the
twistor transform in ten dimensions \cite{lli}.

{}From the other hand the property \p{4} of a superform to close
(called Weil triviality) was considered in the framework of SYM theories
and shown to be essential for deriving consistent chiral anomaly
\cite{wt}.

The other ("geometro-dynamical" \cite{gs2}) term of the twistor-like heterotic
string action alluded above is just needed  to get eq.~\p{5}:
\begin{equation}\label{7}
S_{GD}=\int d^2\xi d^8\eta{\rm P}^q_aE^a_q,
\end{equation}
where ${\rm P}^q_a$ is a Lagrangian multiplier.

Upon getting rid of all auxiliary fields and Lagrange multipliers one
may restore the $N=1$ target-space supersymmetric part of the
conventional heterotic string action. This is possible due to suitable
local symmetries involving transformations of the Lagrange multipliers as
discussed in \cite{dg}. Note that the kinetic and the Wess--Zumino term of
the conventional action follow from the Wess--Zumino--type term \p{1},
while the crucial role of the geometro-dynamical twistor term \p{7} is
to match the dynamics of the string and the worldsheet geometry with the
geometry of the target superspace so that the light-like integrability
takes place (see \cite{n8s,dg} for the details).

To include the chiral fermions into the twistor-like formulation we
propose to add to eq.~\p{1} a term
\begin{equation}\label{8}
S_F=-\int d^2\xi d^8\eta{\rm P}^{\cal{MN}}(ie^+_{\cal
M}\Delta_{\cal N}\Psi^I\Psi^I),
\end{equation}
where the first component $\psi^I\equiv\Psi^I\vert_{\eta=0}$
$(I=1,...,32)$ of the superfield $\Psi^I(\xi,\eta)$ corresponds to the
chiral fermion field with the conformal weight $(-{1\over 2})$ in the
conventional heterotic string approach, and $\Delta_{\cal N}=D_{\cal N}-
E^B_{\cal N}A_B$ is a covariant derivative acting on $\Psi^I(\xi,\eta)$
in the presence of the background gauge superfield $A(X,\Theta)$. Note
that, in particular, the following (anti)commutation relations for
$\Delta_{\cal A}\equiv e_{\cal A}^{\cal N}\Delta_{\cal N}$
\begin{equation}\label{9}
\{\Delta_q,\Delta_r\}=2\delta_{qr}\Delta_-;\qquad [\Delta_q,\Delta_-]=0
\end{equation}
take place on the surface of the background, worldsheet and twistor
constraints, which is again a manifestation of the light-like
integrability.

Similar to eq.~\p{4} the on-shell consistency requires the pull-back
onto the super worldsheet of the form $e^+\wedge\Delta\Psi^I\Psi^I$ to
be closed:
\begin{equation}\label{10}
d(e^+\wedge\Delta\Psi^I\Psi^I)\vert_{M_{ws}}=0.
\end{equation}
This is achieved by taking into account constraints \p{3} and including
into the action a "geometro-dynamical" term for $\Psi^I$:
\begin{equation}\label{11}
S_{FGD}=\int d^2\xi d^8\eta{\rm K}^{rq}\Delta_r\Psi^I\Delta_q\Psi^I,
\end{equation}
where ${\rm K}^{rq}$ is a symmetric Lagrange multiplier superfield.

Varying \p{9} with respect to ${\rm K}^{rq}$ we get
\begin{equation}\label{12}
\Delta_r\Psi^I\Delta_q\Psi^I=0,
\end{equation}
from which, due to the positive definiteness of the quadratic form
\p{12} and in an assumption that the solution of \p{12} is to be
supersymmetric, we have
\begin{equation}\label{13}
\Delta_r\Psi^I(\xi,\eta)=0.
\end{equation}
In particular, in the flat background and in the conformal gauge (when
$\{\Delta_q,\Delta_r\}=2\delta_{qr}\partial_-$) eq.~\p{13} is reduced to
\begin{equation}\label{14}
\partial_-\psi^I=0,
\end{equation}
indicating that the leading component of $\Psi^I(\xi,\eta)$ is indeed
the chiral fermion, while all other components vanish.

Using eqs.~\p{3},\p{9} and \p{12} one may convince oneself that
eq.~\p{10} is valid.

The conventional heterotic fermion action is restored simultaneously
with the $N=1$ supersymmetric part of the heterotic string by gauge
fixing the Lagrange multipliers ${\rm P}^{\cal{MN}}$ to be
\begin{equation}\label{15}
{\rm P}^{+-}=\varepsilon^{+-}\eta^8T,
\end{equation}
with the other components being zero, where T is a string tension
\cite{dg}.

Thus, in particular, in the conformal gauge (and $A_M=0$) we obtain the
standard heterotic fermion action
\begin{equation}\label{16}
S_{F}=T\int d^2\xi i\psi^I\partial_-\psi^I.
\end{equation}

The only thing which remains to check is that there is a local symmetry
of the Siegel type \cite{si} which allows one to gauge away all
propagating degrees of freedom of ${\rm K}^{rq}$. There is indeed such a
symmetry, and the relevant transformations look as follows:
$$
\delta\Psi^I=\epsilon_q\Delta_q\Psi^I,\qquad \delta Q_{\cal
M}=ie^+_{\cal M}\epsilon_q\Delta_q\Psi^I\Psi^I,
\qquad \delta{\rm P}^{\cal {MN}}=0,
$$
$$
\delta{\rm K}^{rq}=-\epsilon^{(r}\left(\Delta_s{\rm K}_s^{q)}+i{\rm
P}^{{\cal M}q)}e^+_{\cal M}\right)
$$
\begin{equation}\label{17}
-{4\over 5}\left[\Delta^{(r}\left(({\rm
K}+i{\rm P}^{{\cal M}-}e^+_{\cal M})\epsilon^{q)}\right)+\delta^{rq}
\Delta_s\left(({\rm
K}+i{\rm P}^{{\cal M}-}e^+_{\cal
M})\epsilon_s\right)\right]+\Delta_s\Lambda^{srq},
\end{equation}
where $\epsilon_q(\xi\eta)$ and $\Lambda^{srq}(\xi,\eta)$ are parameters
of the transformations, the latter parameter being symmetric and
traceless, and ${\rm K}\equiv{1\over 8}\sum_{q}{\rm K}^{qq}$.

The transformations \p{17} (with taking into account \p{9}) eliminate
the traceless part of ${\rm K}^{qr}$ and fix ${\rm K}$ to be a constant.
Hence, ${\rm K}^{qr}$ are completely auxiliary degrees of freedom.

One should notice that the action term \p{11} can be absorbed in the
Wess--Zumino--type term by a suitable shift of $P^{\cal {MN}}$. In that
case the structure of the local symmetries is changed but there are
still enough local transformations (including that similar to
eq.~\p{17}) to gauge fix $P^{\cal {MN}}$ to eq.~\p{15} and a constant.
However, we prefer to keep eq.~\p{11} since then the consideration is
more transparent.

Comming back to eq.~\p{17}, the fact that we gauge fix $K$ to be
a constant seems may cause a problem. Substituting $K=1$ into eq.~\p{11}
retains in the action terms like
$$
\int d^2 \xi
\varepsilon^{q_1...q_8}\psi_{q_1...q_r}\Delta_-\psi_{q_{r+1}...q_8},
$$
where $\psi_{q_1...q_r}=\Delta_{q_1}...\Delta_{q_r}\Psi\vert_{\eta=0}$.
However, eq.~\p{12} still holds in this gauge at $\eta=0$, and eq.~\p{13}
is again valid if the solution is required to be supersymmetric. This
follows from the fact that the highest component of $K$ cannot be gauged
away everywhere. We meet the same situation as in \cite{si}, and  the
arguments proposed by Siegel to show that the constraints are maintained
even when the Lagrange multipliers are gauge fixed are applied here as
well.

In conclusion we have constructed the $n=8$ worldsheet superfield action
for describing heterotic fermions. The action consists of the two terms,
the Wess--Zumino--type term and the geometro-dynamical term, and is
naturally incorporated into the $N=1$, $D=10$ twistor-like heterotic
string action.

When restricted to the solution of eq.~\p{12} that obeys the worldsheet
supersymmetry (eq.~\p{13}) the model reproduces the standard heterotic
string at the classical level. Non supersymmetric solutions of eq.~\p{12}
contain additional propagating fields, which may enter the action with
a kinetic term having the wrong sign, thus causing the problem of
unitarity. A question is whether at the quantum level the physical
sector of the model can be consistently singled out.
So the next step is to perform a
covariant quantization of the theory, which, as one may expect, is to be
a highly nontrivial problem.

\vspace{0.5cm}

\newpage
{\large\bf Acknowledgments}

\medskip
D.S. is grateful to the European Community for financial support under
the contract CEE-SCI-CT92-0789, the Department of Physics of the
University of Padova, and, in particular, the members of the theoretical
group for kind hospitality in Padova.

D.S. is grateful to I. Bandos, S. Krivonos and D. Volkov for stimulating
discussion. M.T. thanks N. Berkovits for clarifying discussion on heterotic
fermions. The authors are grateful to P. Pasti for interest to this work
and useful discussions.

\end{document}